\documentclass[12pt,a4paper]{article}

\usepackage{subfigure}
\usepackage{graphicx}
\usepackage{epsfig}
\date{}
\begin{document}
\title {Hot Nuclear Matter in Asymmetry Chiral 
Sigma Model}
\author {P. K. Sahu$^{1}${\footnote {email:  
pradip@iopb.res.in}, T. K. Jha$^{2}${\footnote {email:tarunjha@iopb.res.in}},
K. C. Panda$^{2}$ and S. K. Patra$^{1}${\footnote {email:
  patra@iopb.res.in}} } \\ 
\\ 
  $^1${\it Institute of Physics, 
 Sachivalaya Marg }\\
{\it Bhubaneswar 751 005, India }
\\
\\$^2${\it P. G. Department of Physics}\\{\it
Sambalpur University}\\
{\it Jyoti Vihar, Burla 768 019, India}}
\maketitle

\begin{abstract}
In the frame work of SU(2) chiral sigma model, the nuclear
matter properties at zero and finite temperature have been investigated.
We have analyzed the nuclear matter equation
of state by varying different parameters, which 
agrees well with the one derived from the heavy-ion 
collision experiment at extreme densities and reliable realistic(DBHF)
model at low density region.
We have then calculated the temperature dependent asymmetric
nuclear matter, also investigated the critical temperature of 
liquid gas phase transition and compared with the experimental
data.
We found that the critical temperature in our model is in the
range of $14-20$ MeV.
\end{abstract}



\newpage
\section{Introduction}
Recently, nuclear equation of state (EOS) is of great interest in
nuclear physics and astrophysics(\cite{ran93} - \cite{zha96}). Specially, in 
the calculation
of nuclear many body problem such as liquid gas phase 
transition\cite{subal00,das92} at 
low density and finite temperature.
The EOS is also useful to study the quark gluon plasma (QGP)
at extreme densities and temperatures(\cite{sahu02a}-\cite{let03}). 
Very recent experiment has confirmed
indirectly more or less the formation of QGP at extreme 
conditions\cite{STAR}.
Also the EOS is a main ingredient to study the evolution of stellar systems, 
and the global properties of neutron star and supernova\cite{sahu00,shen98,brown82}.

To derive the EOS theoretically, many body approaches have been adopted. 
These are Hartree-Fock, Thomas-Fermi and mean-field theory type procedures
\cite{bonche81,serot86}. 
One of them, the relativistic mean-field (RMF) formalism is of great success 
in the theoretical calculation of finite nuclei and infinite nuclear 
matter\cite{serot86,patra01,sahu00,sahu02a}.
The original Walecka RMF model\cite{wal74} has been modified to a great extent 
due to its unrealistic
meson nucleon interactions. For example by adding the non-linear self-interaction 
of scalar mesons in the RMF model\cite{boguta,patra01,sahu02a}, one can describe 
desirable values of saturation 
properties of nuclear matter such as incompressibility, binding energy, saturation 
density and effective nucleon mass. Though the non-linear RMF describe well the 
finite nuclei
and the nuclear matter properties at normal density, it deviates from 
the relativistic Dirac Brueckner
Hartree Fock (DBHF) equation of state\cite{haar87}. 
Because, the DBHF is considered to be the most realistic EOS in the 
non-relativistic
approach at low density realm. Therefore, there are attempts to 
include vector meson self-coupling to reproduce the EOS, compatible to DBHF. 
One can not take arbitrarily the vector-scalar and vector-vector 
interactions for the model to be renormalized. 
However, these interactions can be included in the RMF model, inspired by the 
effective field theory(EFT) approach\cite{fur96}. 

The chiral sigma (CS) model plays significant role in the high density 
matter, because chiral
symmetry is a good hadron symmetry\cite{sahu93}, which is desirable in any
theory of dense hadronic matter. The CS model is analogous to RMF, 
where meson
fields are treated in the mean-field approximation. The beauty of this model 
is that its non-linear
terms simulate the three body forces and is essential to reproduce the nuclear 
matter saturation
properties. A decade ago we proposed a SU(2) CS model, where  the mass of 
the isoscalar vector field
is generated dynamically\cite{sahu93}. The main problem of this theory was 
unrealistic high 
incompressibility. To overcome this shortcoming, recently\cite{sahu00a} we made 
an attempt to introduce the
higher order terms of scalar meson field. In this way, we can reproduce the 
empirical values of
incompressibility, effective mass, binding energy and saturation density. 
Using the same SU(2) CS model
we calculate here the EOS for symmetric and asymmetric nuclear matter at zero 
and finite temperature.
In this calculation, we choose the incompressibility 210, 300 and 380 MeV 
and the effective masses,
0.8, 0.85 and 0.9 of nucleon mass and discuss their applicability to 
the various heavy-ion collision experiments.

The main interest to study finite temperature nuclear EOS is to observe the 
liquid gas phase
transition\cite{li94} near normal nuclear matter density. It is also 
required to estimate 
the critical temperature at liquid gas coexistence point. This feature is 
very much noticeable, because in
the medium energy heavy-ion collisions, one of the theoretical studies of 
dynamics is the liquid gas 
phase transition. Similar feature has been suggested at very low temperature 
in the crust of
neutron star\cite{brown82,peth95}.
In this direction much work has been carried out using both 
non-relativistic \cite{das92}
as well as relativistic \cite{wang00} formalisms. The liquid gas phase 
transition and the critical 
temperature were studied extensively based on the the non-relativistic 
theory(\cite{jaq84}-\cite{latt78}). The estimated critical temperature 
is found to be in the range of 15-20 MeV.

The liquid gas phase also has been studied by many authors based on the 
RMF theory\cite{wang00,hua00}. In the
original Walecka model, the critical temperature in the symmetric nuclear 
matter is 
estimated to be 18.3 MeV\cite{serot86}. This value was brought down to 
14.2 MeV, if the non-linear
terms were included in the model. However, recent experiments in heavy-ion collisions 
for dilute warm nuclear 
matter report a small liquid gas phase region and low critical 
temperature$\sim 13.1\pm0.6$ MeV\cite{li94}.
It has been noticed that the different critical temperature are 
extracted by various
theories. This is because, each theory has its own type of treatments 
of the nuclear 
interactions. Therefore, our motivation is to see such properties 
in the present modified CS sigma model.

The paper is organized as follows: In section II, we present a brief formalism 
of the CS model. In this section we derive the EOS with zero 
and finite temperature for symmetric and asymmetric nuclear matter. 
Results and discussions are displayed in section III. The calculation of the 
EOS for various parameter sets and its sensitivity have been studied and
the liquid gas phase transition has been discussed in this section.
We also compare our results with the recent experimental data. The summary 
and concluding remarks follow in section IV.

\section{The Formalism of SU(2) Chiral Sigma Model}

The {\it SU}(2) chiral sigma Lagrangian can be written as \cite{sahu93,sahu00a}

\begin{eqnarray}
\label{lag}
{\cal L} &=&  \frac{1}{2}\big(
         \partial_{\mu} {\bf \pi} \cdot \partial^{\mu} {\bf \pi}
        +\partial_{\mu} \sigma \partial^{\mu} \sigma
        \big)
- \frac{1}{4} F_{\mu\nu} F_{\mu\nu}
\nonumber \\
&&- \frac{\lambda}{4}\big(x^2 - x^2_o\big)^2
- \frac{\lambda b}{6m^2}\big(x^2 - x^2_o\big)^3
- \frac{\lambda c}{8m^4}\big(x^2 - x^2_o\big)^4
\nonumber \\
&&- g_{\sigma} \bar{\psi} \big(
                \sigma + i\gamma_5 {\bf \tau}\cdot{\bf \pi}
        \big) \psi
+ \bar\psi \big(
i\gamma_{\mu}\partial^{\mu} - g_{\omega}\gamma_{\mu}
\omega^{\mu}\big) \psi
\nonumber \\
&&+ \frac{1}{2}{g_{\omega}}^{2}
x^2 \omega_{\mu}
\omega^{\mu}
+ \frac{1}{24}\xi {g_w}^4(\omega_{\mu}\omega^{\mu})^2 -D\sigma .
\end{eqnarray}
Here $F_{\mu\nu}\equiv\partial_{\mu}\omega_{\nu}-\partial_{\nu}
\omega_{\mu}$ and  $x^2= {\bf \pi}^2+\sigma^{2}$, $\psi$ is the nucleon
isospin doublet, ${\bf \pi}$ is the pseudoscalar-isovector pion field,
$\sigma$ is the scalar field, and $D$ is a constant. 
We work in natural units with $\hbar = c = k_{B} = 1$.

The Lagrangian includes a dynamically generated mass of the 
isoscalar vector field,
$\omega_{\mu}$, that couples to the conserved baryonic current
$j_{\mu}=\bar{\psi}\gamma_{\mu}\psi$.
The constant parameters $b$ and $c$ are included in the higher-order
self-interaction of the scalar field to describe 
the desirable values of nuclear matter properties at saturation
point. Henceforth, we define our model as modified non-linear CS model 
(NCS) in our successive discussions.
In the fourth order term of the omega fields, the quantity $\xi$
is a constant parameter.
Throughout our calculations, for simplicity, we set $\xi$ to zero.
In this model the pion mass $m_{\pi}$ is zero without symmetry
breaking. Thus the last term, $D\sigma$ in the 
Lagrangian is zero in our present calculation.
The interaction of the scalar and the pseudoscalar mesons with the
vector boson generate a mass for the latter
through the spontaneous breaking of the chiral symmetry.
The masses of the nucleon, scalar and vector meson
are respectively given by
\begin{eqnarray}
m = g_{\sigma} x_o,~~ m_{\sigma} = \sqrt{2\lambda} x_o,~~
m_{\omega} = g_{\omega} x_o\ ,
\end{eqnarray}
where $x_o$ is the vacuum expectation value of the $\sigma$ field,
$\lambda~=~({m_{\sigma}}^{2}-{m_{\pi}}^{2})/(2 {f_{\pi}}^{2})$, with
$m_{\pi}$, the pion mass and $f_{\pi}$ the pion decay coupling
constant, and $g_{\omega}$ and $g_{\sigma}$ are the coupling constants
for the vector and scalar fields, respectively.
In the mean-field treatment we ignore the explicit role of $\pi$ mesons.

By adopting mean-field approximation, the equation of motion of fields 
are obtained. This approach has been used extensively to evaluate the 
EOS\cite{sahu02a,patra01,boguta} in any theoretical models for high density matter.
Using the ansatz of the mean-field, 
the equation of motion for the scalar field ($\sigma_0$) in terms
of $m^{\star}/m \equiv x/x_o$ is
 
\begin{eqnarray}
(1-Y^2) -\frac{b}{m^2 c_{\omega}}(1-Y^2)^2
+\frac{c}{m^4c_{\omega}^2}(1-Y^2)^3 \nonumber \\
+\frac{2 c_{\sigma}c_{\omega}n_B^2}{m^2Y^4}
-\frac{2 c_{\sigma}n_S}{m Y}=0\ ,
\label{effmass}
\end{eqnarray}
where $m^{\star} \equiv Ym$ is the effective mass of the nucleon and
$c_\sigma \equiv  g_{\sigma}^2/m_{\sigma}^2 $  and
$  c_{\omega} \equiv g_{\omega}^2/m_{\omega}^2 $ are scalar and vector 
coupling constants respectively. $n_S$ is the scalar density defined in equation (7) 

The equation of motion for the isoscalar vector field is 

\begin{equation}
\omega_0=\frac{ n_B }{g_{\omega} x^2} \ ,
\end{equation}
where in the mean-field limit $\omega$ = $\omega_0$.
The quantity $k_F$ is the Fermi momentum and $\gamma$ is the nucleon spin
degeneracy factor and $n_B$ is the baryon density defined in the next section.

\subsection{Equation of state at zero temperature}

The EOS is calculated from the diagonal components of the
conserved total stress tensor corresponding to the Lagrangian together
with the mean-field equation of motion for the Fermion field and a
mean-field approximation for the meson fields.
The total energy density, $\varepsilon$, and pressure, $P$, of
the many-nucleon system are the following:

\begin{eqnarray}
\label{ep0}
\varepsilon
&=&
          \frac{m^2(1-Y^2)^2}{8c_{\sigma}}
        - \frac{b}{12c_{\omega}c_{\sigma}}(1-Y^2)^3
        + \frac{c}{16m^2c_{\omega}^2c_{\sigma}}(1-Y^2)^4
\nonumber \\
&&+
          \frac{c_{\omega} n_B^2}{2Y^2}
        + \frac{\gamma}{2\pi^2}
                \int _o^{k_F} k^2dk\sqrt{{k}^2 + m^{\star 2}}\ ,
\nonumber\\
P &=&
        - \frac{m^2(1-Y^2)^2}{8c_{\sigma}}
        + \frac{b}{12c_{\omega}c_{\sigma}}(1-Y^2)^3
        - \frac{c}{16m^2c_{\omega}^2c_{\sigma}}(1-Y^2)^4
\nonumber \\
&&+
          \frac{c_{\omega}n_B^2}{2Y^2}
        + \frac{\gamma }{6\pi^2}
                \int _o^{k_F} \frac{k^4dk}{\sqrt{{k}^2 + m^{\star 2}}}\ .
\end{eqnarray}
The energy per nucleon is $E/A=\varepsilon/n_B$,
where $\gamma=4$ for symmetric nuclear matter and $\gamma=2$ for neutron matter.

The baryon density $n_B$ and scalar density $n_S$ are defined as
\begin{equation}
n_B= \frac{\gamma}{(2\pi)^3}\int^{k_F}_o d^3k ,
\end{equation}
and 
\begin{equation}
n_S= \frac{\gamma}{(2\pi)^3}\int^{k_F}_o\frac{m^* d^3k}
         {\sqrt{k^2+m^{\star 2}}},
\end{equation}
respectively, which are used in eq.(\ref{effmass}).

\subsection{Equation of state at finite temperature}

The EOS for finite temperature can be defined in the same 
manner as zero temperature, these are as follows:

\begin{eqnarray}
\label{ept}
\varepsilon(T)
&=&
          \frac{m^2(1-Y^2)^2}{8c_{\sigma}}
        - \frac{b}{12c_{\omega}c_{\sigma}}(1-Y^2)^3
        + \frac{c}{16m^2c_{\omega}^2c_{\sigma}}(1-Y^2)^4
\nonumber \\
&&+
          \frac{c_{\omega} n_B^2}{2Y^2}
        + \frac{\gamma}{2\pi^2}
                \int _o^{\infty} k^2dk\sqrt{{k}^2 + m^{\star 2}}
        (f(T)+\bar f(T))\ ,
\nonumber\\
P(T) &=&
        - \frac{m^2(1-Y^2)^2}{8c_{\sigma}}
        + \frac{b}{12c_{\omega}c_{\sigma}}(1-Y^2)^3
        - \frac{c}{16m^2c_{\omega}^2c_{\sigma}}(1-Y^2)^4
\nonumber \\
&&+
          \frac{c_{\omega}n_B^2}{2Y^2}
        + \frac{\gamma }{6\pi^2}
                \int _o^{\infty} \frac{k^4dk}{\sqrt{{k}^2 + m^{\star 2}}}
         (f(T)+\bar f(T))\ .
\end{eqnarray}
The baryon and scalar density at finite temperature are respectively
modified as
\begin{equation}
\label{nb}
n_B(T)= \frac{\gamma}{(2\pi)^3}\int^{\infty}_o d^3k (f(T)-\bar f(T)) ,
\end{equation}
and 
\begin{equation}
\label{ns}
n_S(T)= \frac{\gamma}{(2\pi)^3}\int^{\infty}_o\frac{m^* d^3k}
        {\sqrt{k^2+m^{\star 2}}} (f(T) + \bar f(T)) .
\end{equation}
The nucleon and anti-nucleon distribution functions $f(T)$ and $\bar f(T)$,
are respectively, expressed as

\begin{equation}
f(T)=\frac{1}{\exp{[(E^{\star}+\nu)/T]}+1}
\end{equation}
and
\begin{equation}
\bar f(T)=\frac{1}{\exp{[(E^{\star}-\nu)/T]}+1} .
\end{equation}
where $E^{\star} = \sqrt{k^2+m^{\star ^2}}$, $T$ is temperature and 
the chemical potential $\nu = \mu - g_w w_o$. These distribution functions
are used in  eq.(\ref{ept}-\ref{ns}).

\subsection{Asymmetric nuclear matter}
 
For asymmetric matter, the extra contribution to Lagrangian eq.(\ref{lag}) 
due to the interaction of the isospin triplet $\rho-$meson is given  
as 

\begin{eqnarray}
\label{rho}
\nonumber \\
&&-\frac {1}{4}{\bf G}_{\mu\nu}\cdot{\bf G}^{\mu\nu}
+\frac{1}{2}m^2_{\rho}{\bf \rho}_{\mu}\cdot{\bf \rho}^{\mu}
-\frac{1}{2}g_{\rho}\bar\psi
                ({\bf \rho}_{\mu}\cdot{\bf \tau}\gamma^{\mu})
                \psi\ .
\end{eqnarray}
where ${\bf G}_{\mu\nu} \equiv \partial_{\mu}{\bf \rho}_{\nu}-\partial_{\nu}
{\bf \rho}_{\mu}$. This term accounts for asymmetric nuclear matter with 
mixture of protons and neutrons only.

Similarly, the equation of motion for $\rho-$meson is obtained from eq.(\ref{rho})
as:

\begin{equation}
\rho^3_o = \frac{g_{\rho}}{2m_\rho^2} (n_p-n_n)\ ,
\end{equation}
where $n_p$ and $n_n$ are number density of proton and neutron, respectively.
The inclusion of the $\rho-$meson in the Lagrangian will contribute the term

\begin{equation}
+ ~ m^2_{\rho}(\rho_o^3)^2/2
\end{equation}
to the energy density and pressure as given above eqs.(\ref{ep0},\ref{ept}) 
for both zero and finite temperature.
From the semi-empirical nuclear mass formula, the symmetric energy
coefficient is
\begin{equation}
a_{\rm sym} = \frac{c_{\rho} k_F^3}{12\pi^2} + \frac{k_F^2}{6\sqrt{(k_F^2
+M^{\star 2})}}\ ,
\end{equation}
where $c_{\rho} \equiv g^2_\rho/m^2_{\rho}$ and $k_F=(6\pi^2n_B/\gamma)
^{1/3} (n_B=n_p+n_n)$.
We fix the coupling constant $c_{\rho}$ by requiring that $a_{\rm sym}$
correspond to the empirical value, 32 $\pm$ 6 MeV\cite{moll88}.
This gives $c_{\rho}=4.66~ \hbox{fm}^2$ for $a_{\rm sym}$=32 MeV.
The chemical potential is redefined as $\nu = \mu - g_w w_o +\tau^3 g_{\rho}
\rho^3_{0}$, due to presence of asymmetric nuclear matter, where $\tau^3$
is $+1/2$ for neutron and $-1/2$ for proton.

Also we introduce the asymmetric parameter, $\alpha$ to describe the asymmetric
nuclear matter. This is defined as 
\begin{equation}
\label{alpha}
\alpha=\frac{n_n-n_p}{n_n+n_p},
\end{equation}
where $\alpha$=0 for symmetric matter nuclear matter ($\gamma=4$) and $\alpha$=1 for the
pure neutron matter ($\gamma=2$).

\begin{figure}[ht]
\epsfxsize=10cm
\begin{center}
\includegraphics[width=10cm,height=12cm,angle=-90]{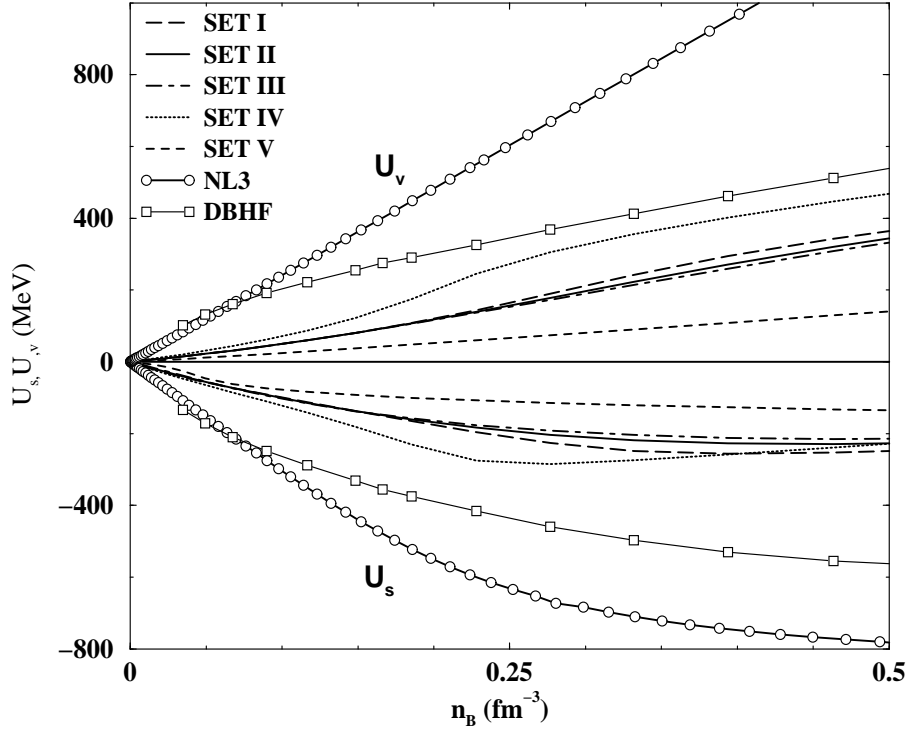}
\end{center}
\caption{ The scalar and vector potentials for various parameter sets of
NCS model with baryon density. The DBHF result is from the Bonn A potential 
and NL3 parameter set is from the relativistic mean-field theory.
}\label{poten}
\end{figure}

\section{Results and discussions}

In the EOS eqs.(\ref{ep0}-\ref{ept})
for both zero and finite temperature, the four parameters are:
the nucleon coupling to the scalar and the vector fields,
$c_{\sigma}$ and $c_{\omega}$,
and the coefficients in the scalar potential terms, $b$ and $c$.
These are obtained by fitting at the saturation point: the binding
energy/nucleon ($B/A~=~ -16.3$ MeV), baryon density ($n_0~ =~ 0.153~\hbox{fm}^{-3}$),
incompressibility ($K$ = 300 MeV) and effective (Landau) mass 
($m^{\star}~=~0.85M$)\cite{moll88}.
In our calculation we have chosen the effective mass from $0.8-0.9M$,
to observe the sensitivity of EOS at high density region. Another 
interesting point we note that by changing the effective mass,
the EOS can be compared well with the recent one which has been extracted 
from the heavy-ion collisions data\cite{data02}. We will discuss this below.
The nuclear incompressibility is somewhat uncertain at saturation
and therefore we take in the range of $210-380$ MeV. The desirable 
values of effective mass and nuclear matter incompressibility are chosen 
in accordance with recent heavy-ion collision data\cite{sahu02,data02}.
These parameters are listed in Table I.
\begin{table}
\caption{Different parameter sets for the NCS model.}
\vskip 0.1 in
\begin{center}
\begin{tabular}{cccccccccccc}
\hline
\hline
\multicolumn{1}{c}{set}&
\multicolumn{1}{c}{$c_{\sigma}$}&
\multicolumn{1}{c}{$c_{\omega}$} &
\multicolumn{1}{c}{$b/m^2$} &
\multicolumn{1}{c}{$c/m^4$} &
\multicolumn{1}{c}{$K$} &
\multicolumn{1}{c}{$m^{\star}/m$} \\
\multicolumn{1}{c}{ } &
\multicolumn{1}{c}{($fm^2$)} &
\multicolumn{1}{c}{($fm^2$)} &
\multicolumn{1}{c}{($fm^2$)} &
\multicolumn{1}{c}{($fm^4$)}&
\multicolumn{1}{c}{($MeV$)} &
\multicolumn{1}{c}{}\\
\hline
I &8.86&1.99&-12.24&-31.59&210&0.85 \\
II&6.79&1.99&-4.32&0.165&300&0.85 \\
III&5.36&1.99&1.13&22.01&380&0.85 \\
IV&8.5&2.71&-9.26&-40.73&300&0.8 \\
V&2.33&1.04&9.59&46.99&300&0.9 \\
\hline
\end{tabular}
\end{center}
\end{table}

\begin{figure}[ht]
\epsfxsize=10cm
\begin{center}
\includegraphics[width=10cm,height=12cm,angle=-90]{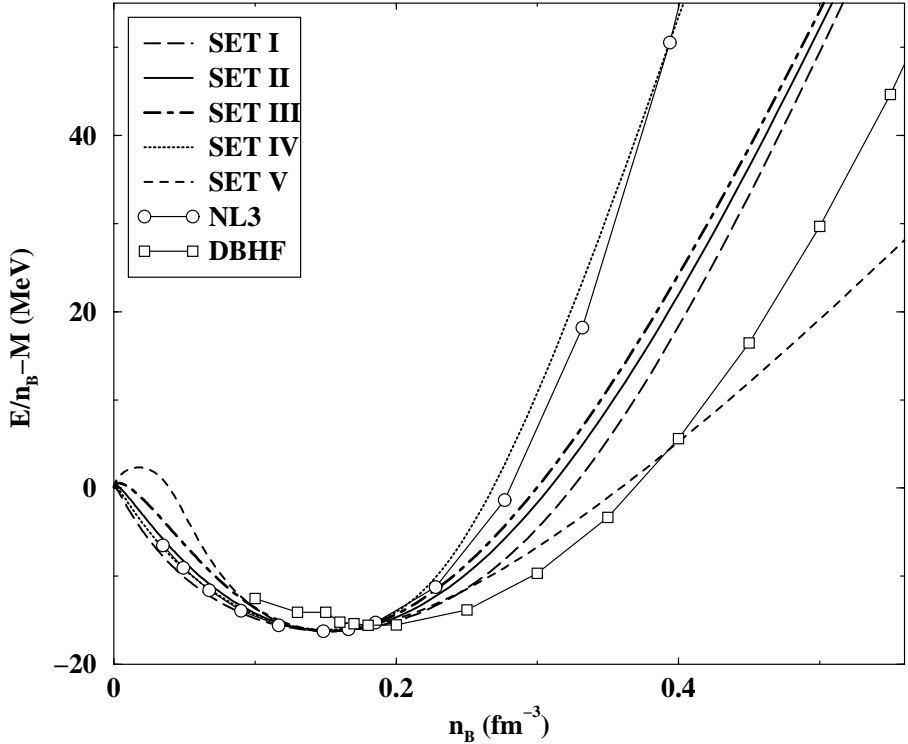}
\end{center}
\caption{Same as figure\ref{poten}, but for energy per particle
}\label{eos}
\end{figure}

\subsection{At zero temperature limit}
The scalar $U_s~(=~g_\sigma \sigma_0)$ and vector $U_v~(=~g_\omega \omega_0)$ 
potentials versus baryon density are displayed in Fig. 1 for
five parameter sets as listed in Table I. We compare these potentials with
the more realistic Dirac--Brueckner--Hartree--Fock (DBHF) (Bonn-A parametrization) 
\cite{li92} and the 
standard $\sigma-\omega$ non-linear relativistic mean-field (NL3 parameter set) 
\cite{lala97} potential, that are available in literature.
It is shown in Refs.(\cite{Gm91}-\cite{Gm92a}) that the
standard $\sigma-\omega$ model with scalar self-couplings describes the
saturation point and the data for finite nuclei successfully, do not follow
the trends of the DBHF properly. In the RMF model, the vector potential 
increases linearly with density and gets stronger as it does not depend on the 
non-linear terms of the vector meson. However, in DBHF
it bends down (see Fig. 1), because it has density dependent potentials. 
The scalar potential overestimates the DBHF result at high density in order to
compensate for the strong repulsion in the vector channel. The  $U_s$ and $U_v$
results obtained by NCS model are quite low at near and below the nuclear
matter saturation density. At high density, the result obtained by parameter set IV
is comparable with the DBHF calculation, whereas other sets underestimate the
results of both NL3 and DBHF models. In the scalar case, all the parameter sets
of NCS model give a low value due to the strong scalar coupling. The smaller value 
of $U_v$ is counter balanced by the higher $U_s$ and gives a similar
magnitude of the total potential, compared to DBHF. For example, at $\sim 3n_0$ the
total potential is $\sim -58$ MeV for DBHF, whereas this is $\sim -73$ MeV in
set II of NCS model. Similarly, for all other sets this varies from $-17$ to 
$-200$ MeV for NCS model and for NL3, it is $-327$ MeV. This feature reflects 
in the EOS (discussed in next figure).

Now we compare the EOS of our calculations with the NL3 and DBHF models in Fig. 2.
Here we find that sets IV  and V match with NL3 and DBHF, respectively 
up to three times the nuclear matter density. It is to be noted that the 
EOS obtained by DBHF is trusted upto two times of nuclear matter
density\cite{Su94,patra01a}. The difference between the two sets IV (stiff) and V (soft)
is only due to the different effective mass for the same 
incompressibility ($K$ = 300 MeV). 
The all other three
sets namely, I (K=210 MeV), II (K=300 MeV) and III (K=380 MeV) 
are having same effective mass with different incompressibilities. 
From this graph, we note that the stiffness or softness of EOS is
insignificant with incompressibilities in comparison to NL3 and DBHF. 

\begin{figure}[ht]
\epsfxsize=10cm
\begin{center}
\includegraphics[width=10cm,height=12cm,angle=-90]{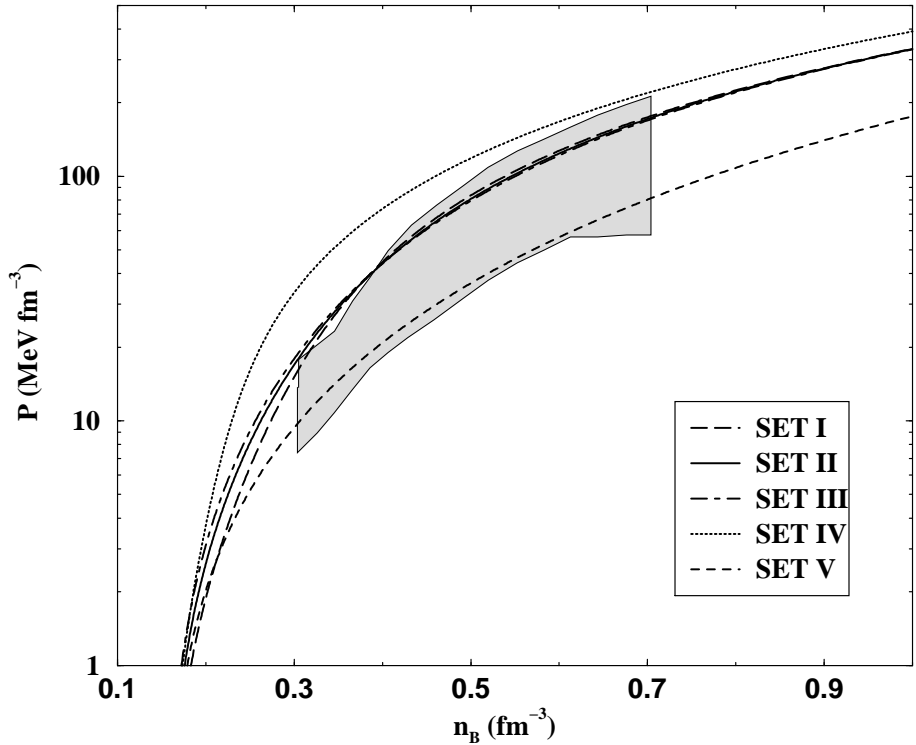}
\end{center}
\caption{The pressure versus baryon density for different sets of NCS model. The
shaded block represents the EOS consistent with  experimental data \cite{data02}.
}\label{press-den}
\end{figure}

In Fig. 3 we compare our EOS with the predicted experimental values obtained from
the heavy-ion collisions data\cite{data02}. The overall EOS are 
good fit to experimental data. If we consider more vividly, then we notice 
that the EOS having incompressibility, $K$=300 MeV and $m^{\star}/m$=0.9 (set
V) fits well. However, $K$ =210 (set I) and 300 (set II) MeV with 
$m^{\star}/m$=0.85 also agree, but slightly deviate from the data at low 
density. In addition, $K$=300 MeV 
and $m^{\star}/m$=0.8 (set IV) shows more stiffer EOS. In general, 
the set II ($K$=300 MeV and $m^{\star}/m$=0.85) explains EOS fairly well and hence could
be the ideal parameterization (set II). Note that
the value for EOS predicted by experiment may change due to the momentum dependent 
potential as given in Ref.\cite{sahu02}. We recall here that all the
sets considered in the present calculations (set I$-$V) compare
well with the DBHF prediction at low density (see Fig. 2).
Whereas this figure represents a clear picture of the EOS at high 
densities.

\begin{figure}[ht]
\epsfxsize=10cm
\begin{center}
\includegraphics[width=10cm,height=12cm,angle=-90]{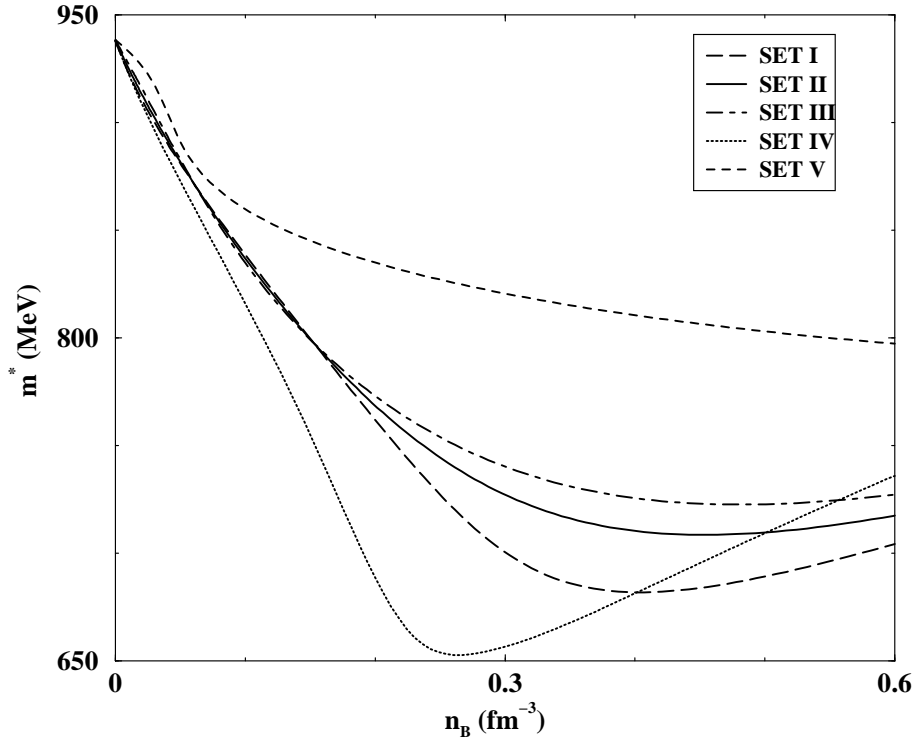}
\end{center}
\caption{ For different five parameter sets of NCS model, the effective
mass is shown with baryon density. 
}\label{effective}
\end{figure}

For the sake of completeness, the effective masses for various sets as a function 
of baryon density are compared in Fig. 4. The features are similar to those 
as given in Fig. 2. That means, sets IV and V give drastically different
curves owing to different effective masses at saturation density for same
incompressibility, as expected. However, we get marginal changes for
other three sets I$-$III beyond nuclear matter density, which represent
the different incompressibilities. It is interesting
to note that $m^{\star}$ for set IV again increases with density
for $n_B\sim 2n_0$ due to strong repulsive force. The effective mass of
set V decreases monotonically (see Fig. 4), which has a strong attractive
force. 

\subsection{At finite temperature limit}

\begin{figure}[ht]
\epsfxsize=7cm
\begin{minipage}{0.4\textwidth}
\includegraphics[width=7cm,height=7.5cm,angle=-90]{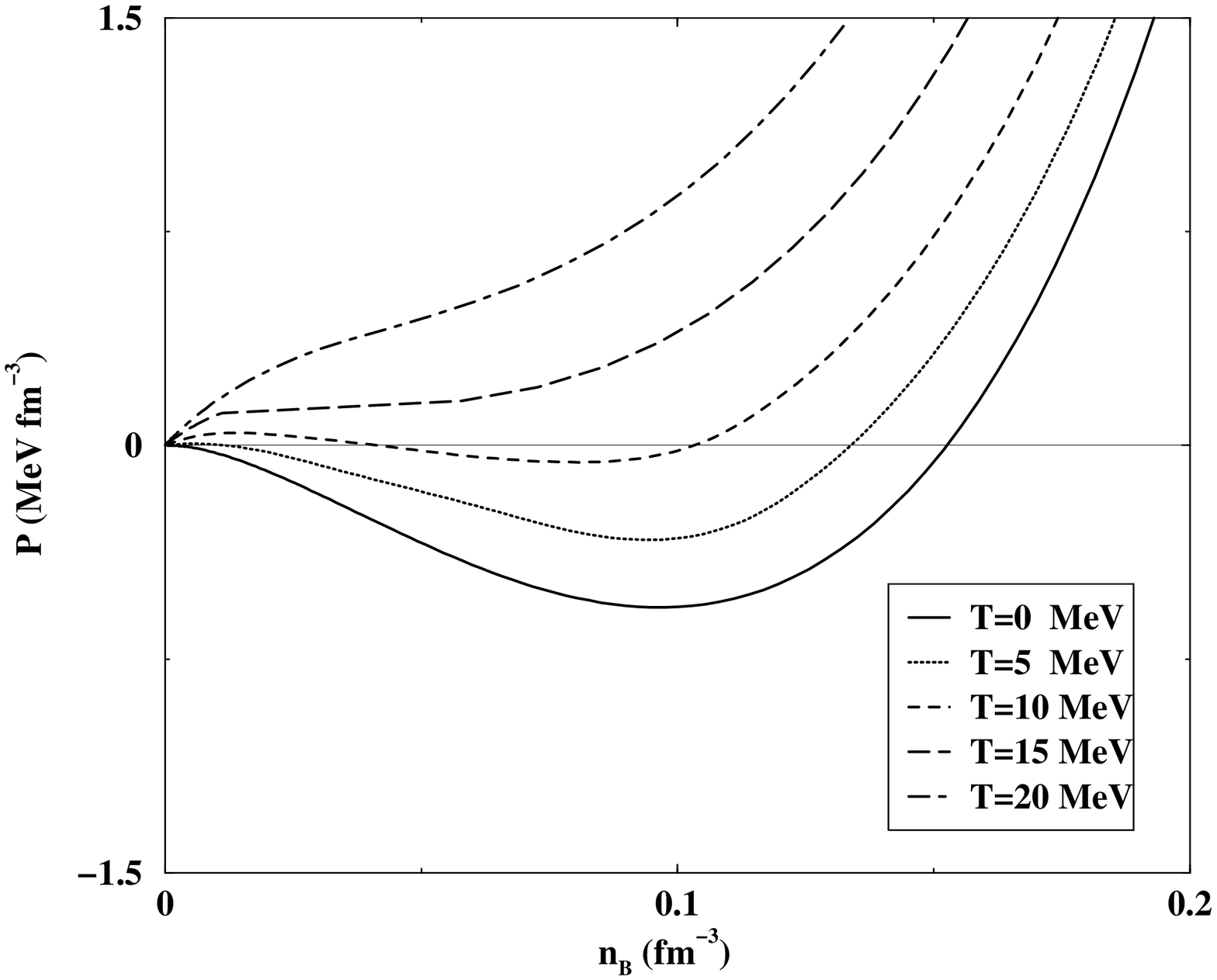}
\end{minipage} \
\begin{minipage}{0.4\textwidth}
\hspace*{1.5cm}
\includegraphics[width=7cm,height=7.5cm,angle=-90]{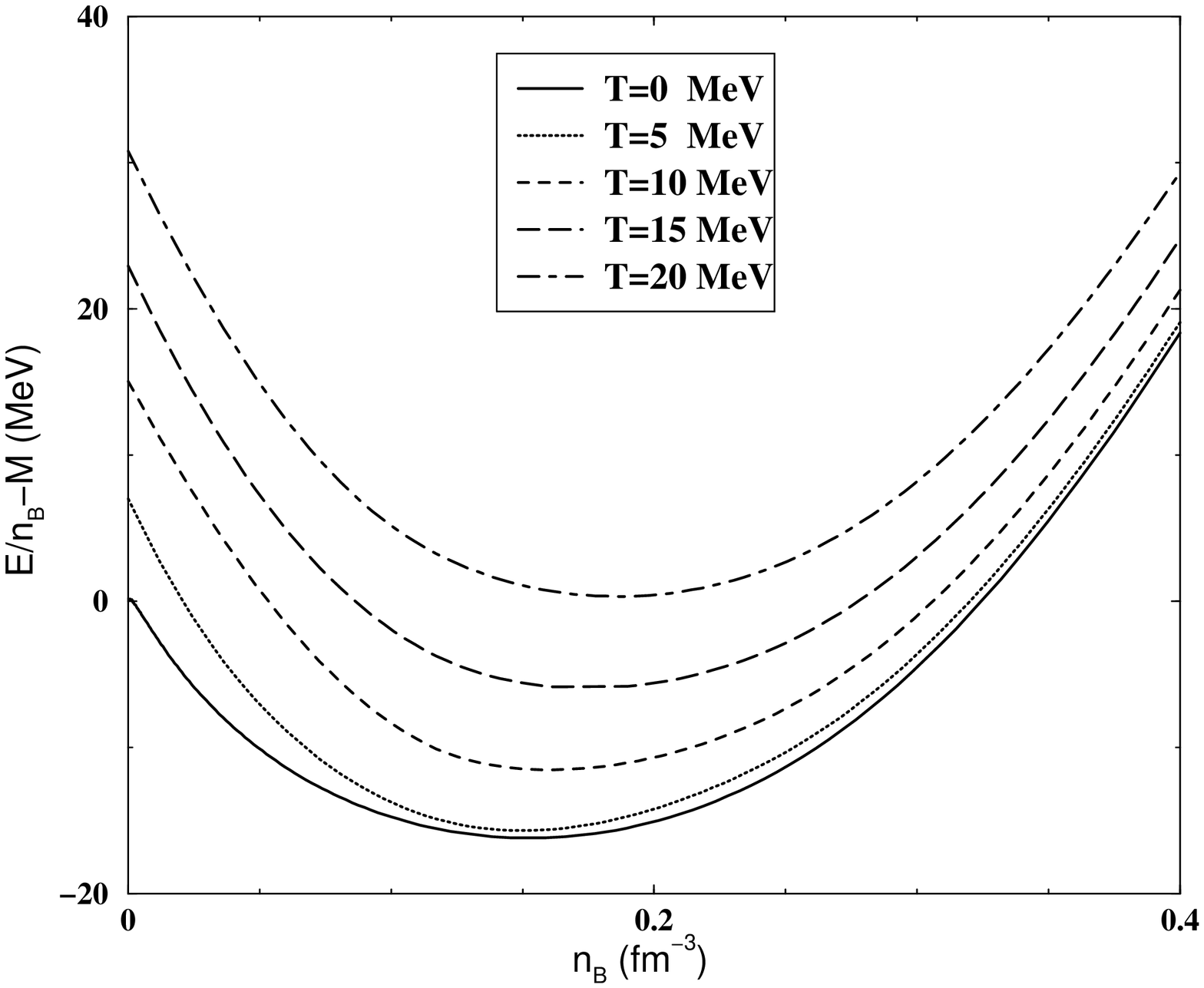}
\end{minipage}
\caption{The left panel of figure represents pressure versus baryon density and the 
right panel shows the energy per particle as a function of baryon density 
for the set I for different temperature.}\label{setI}
\end{figure}

\begin{figure}[ht]
\epsfxsize=7cm
\begin{minipage}{0.4\textwidth}
\includegraphics[width=7cm,height=7.5cm,angle=-90]{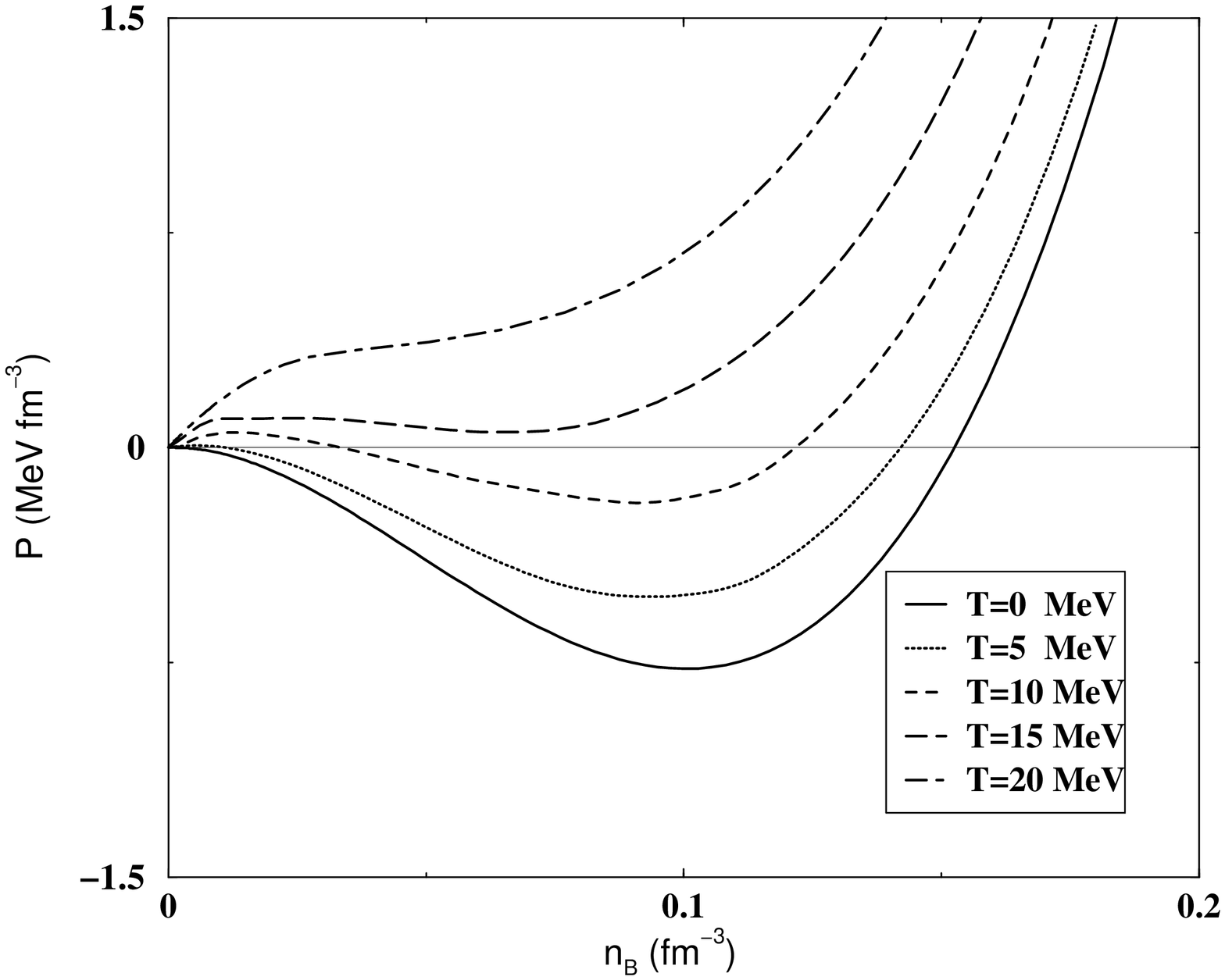}
\end{minipage} \
\begin{minipage}{0.4\textwidth}
\hspace*{1.5cm}
\includegraphics[width=7cm,height=7.5cm,angle=-90]{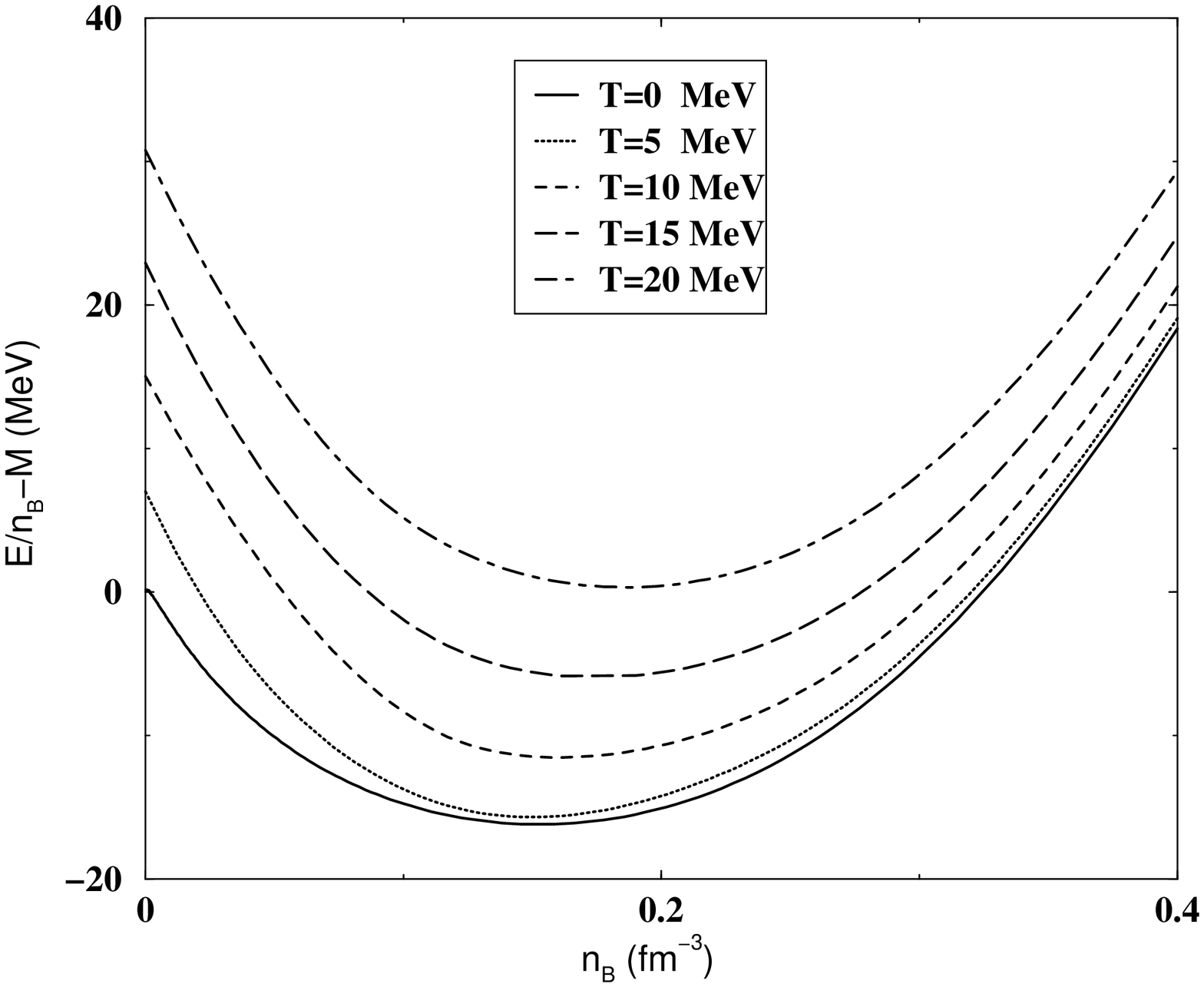}
\end{minipage}
\caption{Same as figure \ref{setI} but for set II.}\label{setII}
\end{figure}

\begin{figure}[ht]
\epsfxsize=7cm
\begin{minipage}{0.4\textwidth}
\includegraphics[width=7cm,height=7.5cm,angle=-90]{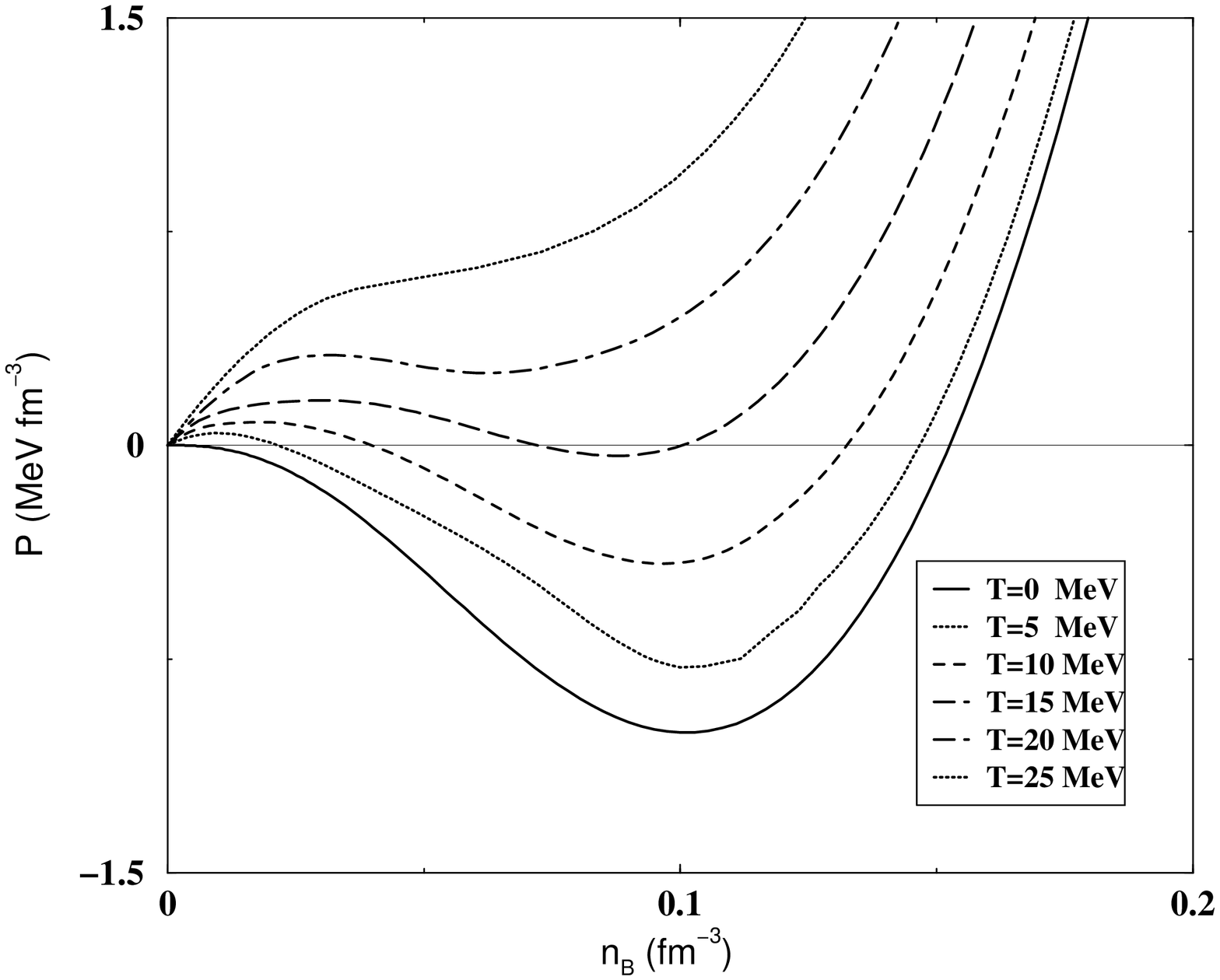}
\end{minipage} \
\begin{minipage}{0.4\textwidth}
\hspace*{1.5cm}
\includegraphics[width=7cm,height=7.5cm,angle=-90]{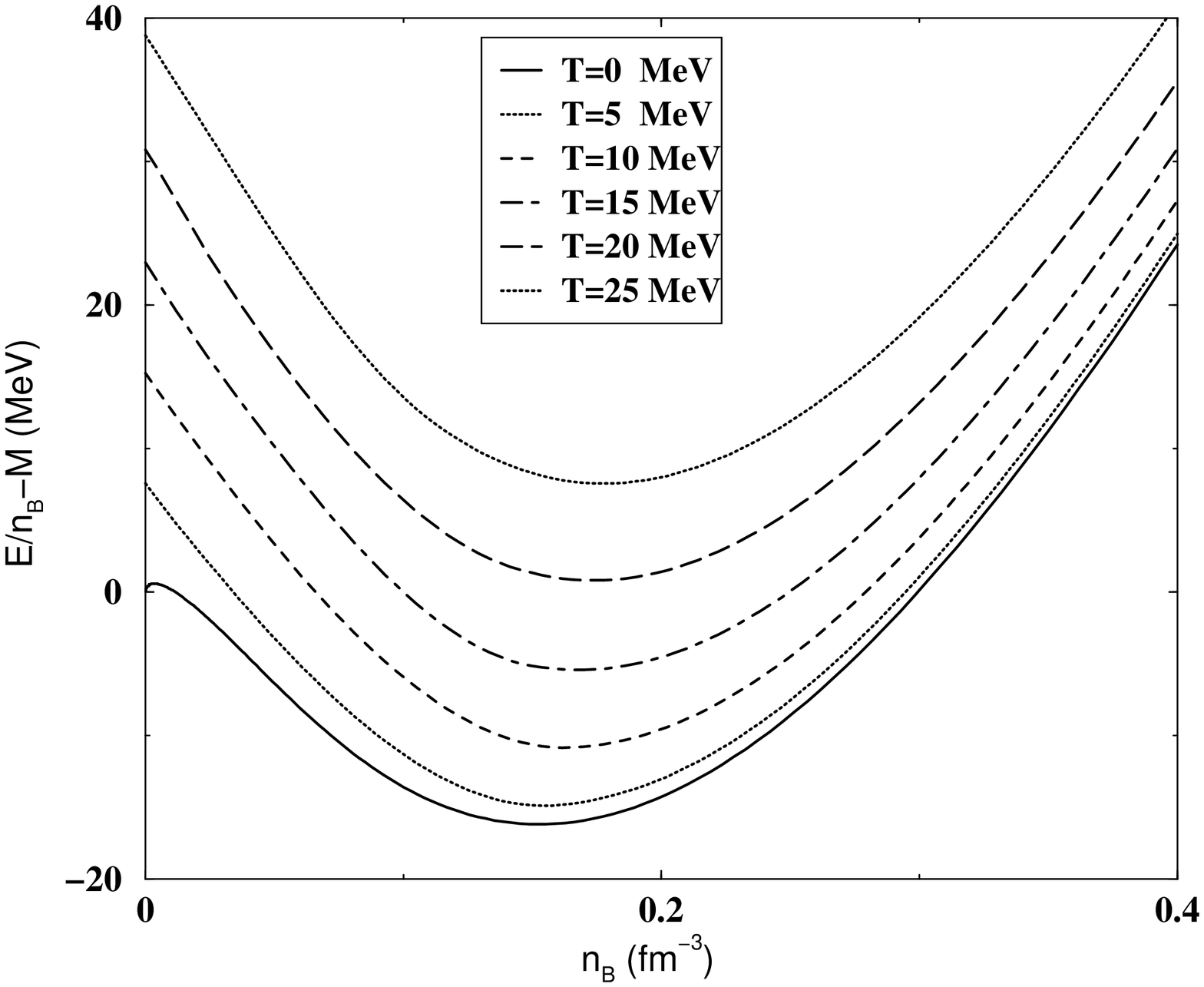}
\end{minipage}
\caption{Same as figure \ref{setI} but for set III}\label{setIII}
\end{figure}

Here we analyze the effect of temperature on the EOS, explicitly near
the nuclear matter density, such as liquid gas phase transition.  
The pressure versus baryon density is plotted to show the liquid gas phase 
transition for various temperature ranging from 0 to 20 MeV in left
panel of Fig. 5(a). In this figure 
we have taken the parameter set I for symmetric matter. At zero temperature, 
there is a nice pocket (isotherm), which means that the liquid and gas phase 
are well separated with each other by an unphysical region, where
the pressure is negative.  One can make a smooth transition from
liquid to gas state by making a Maxwell construction\cite{serot95}.
This pocket gradually decreases with increasing temperature. At a 
particular temperature, the pocket vanishes and is marked as the 
pure gas state. At this point the pressure gradient with respect to
density (inflection point) is zero ($\partial P/\partial n_B|_{T_c}= 
\partial^2 P/\partial {n_B}^2|_{T_c}=0$) and is noted as the critical point 
for liquid gas phase transition. In other words, the point where the 
two phases can not be distinguished from each other for a particular 
temperature is called the critical point. In this figure, 
the critical temperature $T_c=$14.2 MeV which corresponds to 
pressure $P_c$=0.14 MeV/fm$^3$ and density $n_c$=0.035 fm$^{-3}$.

The critical temperature obtained by the density dependent relativistic 
mean-field theory
\cite{hua00} ($T_c=$12.66 MeV) and the experimental value ($T_c=13.1\pm0.6$ MeV)
\cite{li94} are comparable to the $T_c$ obtained in our calculation
with set I. As we emphasized in section I, if one consider 
the original Walecka model \cite{serot86} (no non-linear terms),
the critical temperature is  $T_c\approx 18.3$. This can be reduced to 
$T_c\approx 14.2$ MeV \cite{fur97}, when one introduces the non-linear terms in
the scalar field.
The derivative scalar coupling model \cite{mal98} gives a low critical
temperature ranging from $T_c=13.6-16.5$ MeV depending on the parameter sets.
The critical temperature extracted by DBHF approach is 15.0 MeV \cite{haar87}.
Also, it is reported by Baldo et al \cite{baldo95} that a very low $T_c$ of about
$8-9$ MeV is obtained in relativistic Dirac-Brueckner calculation.
Therefore, it can be concluded from the above models that the critical temperature 
varies from $8-19$ MeV depending on the formalisms and the parameters used. In our
present investigation, we also find a large range of $T_c$ from 14 $-$ 20 MeV
depending on the parameter sets, which will be discussed below. Thus our model
is compatible with the other relativistic and non-relativistic models.

The right panel of Fig. 5(b) displays the energy per particle versus
baryon density for the symmetric matter with set I. With increasing temperature, 
the system becomes less bound in comparison to zero temperature. 
At $T=T_c$, the energy per particle is $-2.04$ MeV.

In Figs. 6(a) (set II) and 7(a) (set III), we show the similar plot like Fig. 5(a)
for symmetric matter. 
The $T_c$ obtained are 16.8 MeV and 20.4  MeV. The corresponding
$P_c$ are 0.22 and 0.36  MeV/fm$^3$ and $n_c$ are 0.044 and 0.051 fm$^{-3}$, 
respectively for set II and III. 
From these values we observe that
the critical temperature $T_c$ increases with incompressibility. The critical
point shifts toward lower density and pressure for softer EOS.
These are comparable to other relativistic\cite{serot95} 
and non-relativistic models\cite{jaq84}.
Similarly, the energy per particle are presented in Figs. 6(b) and 7(b) for 
the sets II and III, at $T=T_c$, the energy per particle are 4.8
and 13.3 MeV, respectively. From the above figures (Figs. 5(b)-7(b)), one
may notice that the system becomes less bound with increasing incompressibility.

\begin{figure}[ht]
\epsfxsize=7cm
\begin{minipage}{0.4\textwidth}
\includegraphics[width=7cm,height=7.5cm,angle=-90]{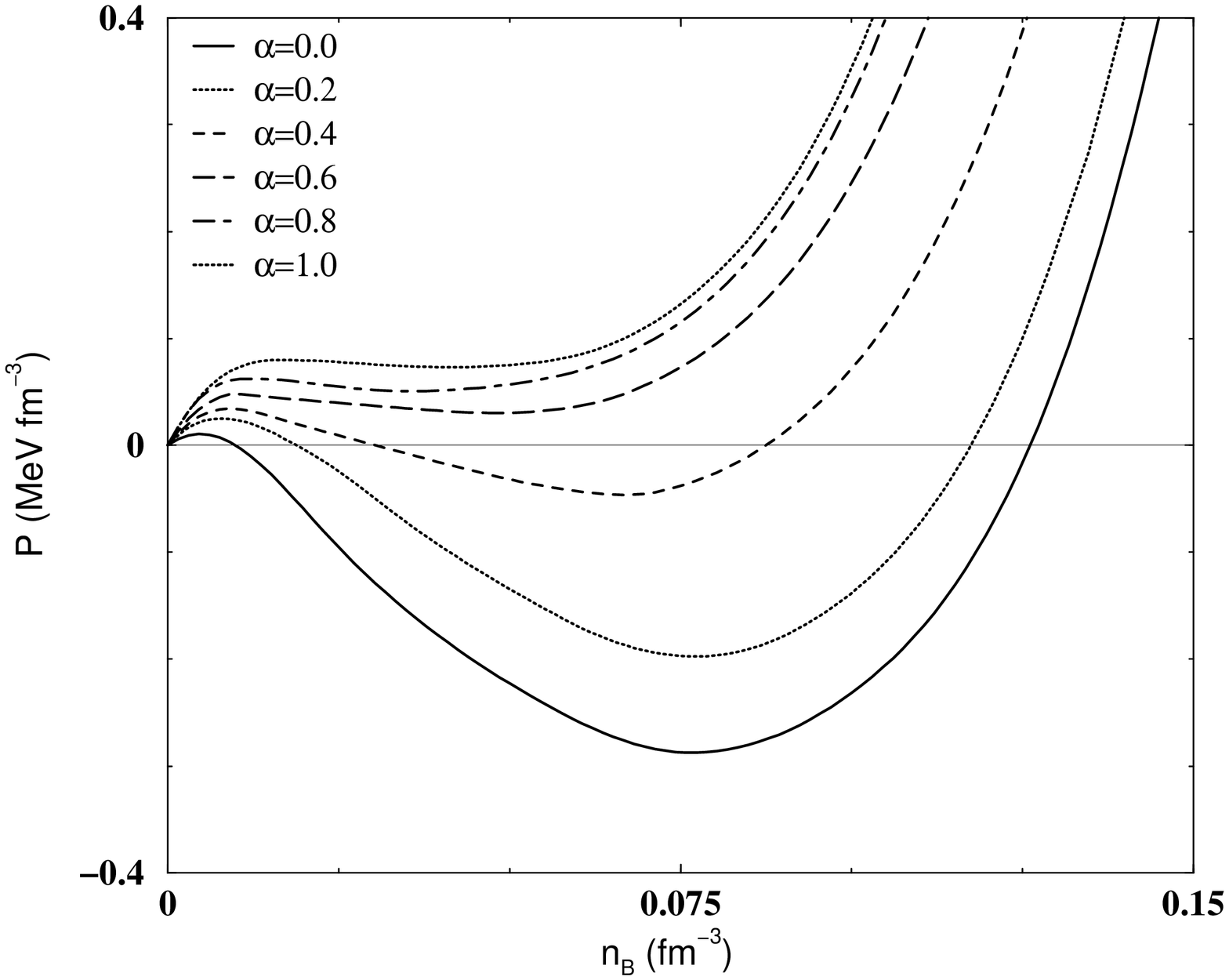}
\end{minipage} \
\begin{minipage}{0.4\textwidth}
\hspace*{1.5cm}
\includegraphics[width=7cm,height=7.5cm,angle=-90]{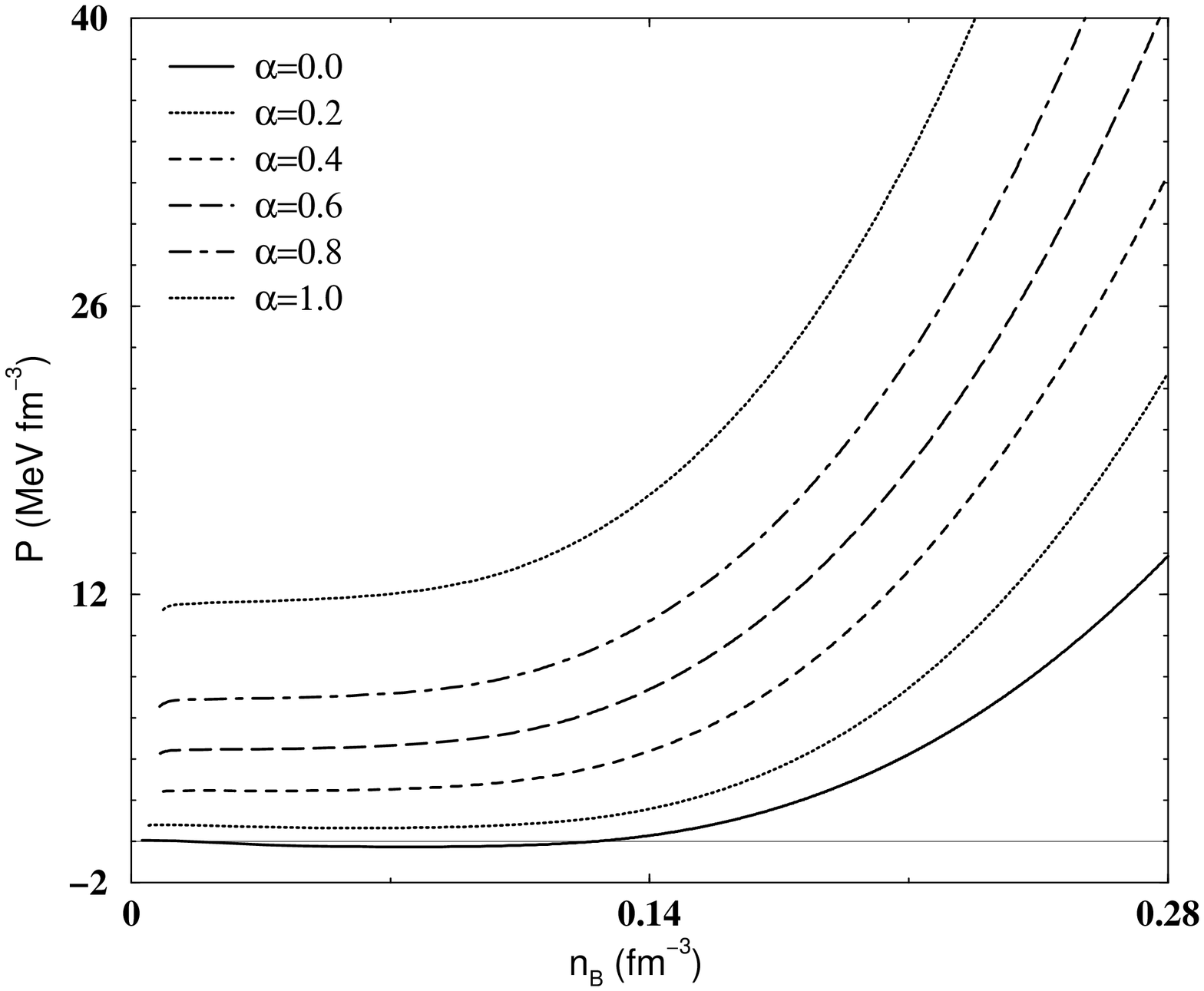}
\end{minipage}
\caption{The pressure as a function of the baryon density for fixed $T$=10 
MeV with various asymmetric parameter $\alpha$. The left panel is without
$\rho-$meson and the right panel is with $\rho-$meson inclusion for set II.
}\label{alpha}
\end{figure}

As we mentioned earlier, sets I and II give reasonably good  fit to the
DBHF and experimental data (see Figs. 2 and 3), we consider hereafter set II
for the rest of our discussions. The reason to choose set II is that it is
compatible with the description of heavy-ion collision data \cite{sahu02}. 
In our further discussions, we study the asymmetric, $\alpha$ dependence 
of the system for a fixed temperature, say for example  $T=10$ MeV. Moreover in this section,
we analyzed the effect of $\rho-$meson on the nuclear system. Also,
the behaviour of effective mass with temperature and the critical temperature
$T_c$ as a function of asymmetric parameter $\alpha$ are studied. 

In Fig. 8(a), the pressure versus baryon density for different 
$\alpha$ at a fixed temperature $T$=10 MeV using parameter set II is displayed.
In this case, we have not included the effect of $\rho-$meson. 
The liquid gas phase
transition disappears at $\alpha > \sim 0.6$, below which
the pressure shows a minimum with respect to density, that means 
there is a phase boundary between two phases as shown in Fig. 8(a).
Also for different 
$\alpha$, we plot pressure versus baryon density for fixed 
$T$=10 MeV with inclusion of $\rho-$mesons in the nuclear matter 
for parameter set II. The graph (Fig. 8(b)) looks very 
similar to Fig. 8(a), but the pressure rapidly increases as the 
extra repulsive force generated from the $\rho-$mesons and hence, the 
liquid gas phase transition vanishes at $\alpha > \sim 0.2$.

The asymmetric parameter, $\alpha$ versus $T_c$ is displayed in Fig. 9 for 
the set II without considering $\rho-$meson. The value of critical 
temperature $T_c$ reduces from 
nuclear matter $\alpha$=0 ($T_c$=16.8 MeV) to neutron matter $\alpha$=1 ($T_c$
=11.2 MeV). It shows that the liquid gas phase transition is more probable
in neutron matter than the pure symmetric nuclear matter. The similar
behaviour has been reported in Ref.\cite{wang00} within the effective 
nuclear model based on the mean-field approximation. That means the 
pressure generated from repulsive saturation force plays vital role to 
undergo early phase transition.

The effect of high temperature on the effective mass, which play a dominant role
in the EOS (eq.\ref{ept}) is shown in Fig. 10. Also EOS at high temperature is useful 
to study the supernova simulation, such as the mechanism and whole phenomena
of supernova explosion\cite{lat91}
In the left panel of Fig. 10 (10(a)), we show the variation of 
$m^*$ with $n_B$ at $T~=~25$, 50 and 100 MeV in set II. The trend of
the curves up to temperature $T=$100 remains similar. It is clear from the figure
that $m^*$ increases gradually with $T$. 
This effect is attributed to the pair formation due
to anti-particle production. For example, the change of effective mass from zero to three
times nuclear matter density, is around $25\%$ of nucleon mass upto $T=100$ MeV. 
In the right panel of Fig. 10 (10(b)), the pressure and energy are plotted for different
temperatures, which are function of baryon density. From this figure, we observe that
the EOS gets stiffer with increasing temperature. The reason of stiffness is that 
the extra thermal energy and pressure contribution comes from the anti-baryons.

\begin{figure}[ht]
\epsfxsize=10cm
\begin{center}
\includegraphics[width=10cm,height=12cm,angle=-90]{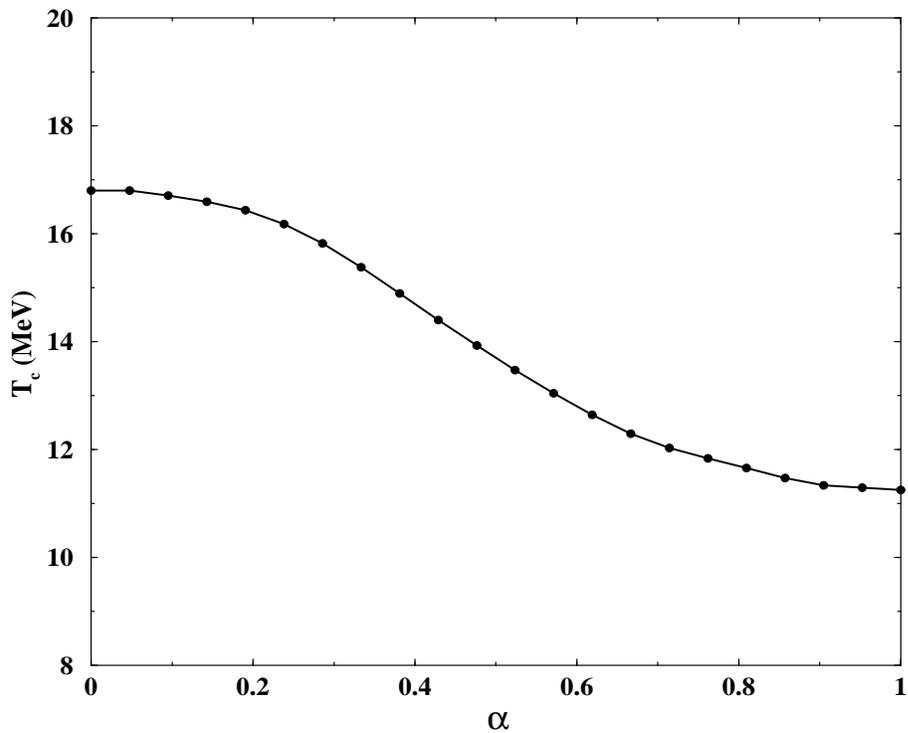}
\end{center}
\caption{ The $T_c$ as a function of asymmetric parameter $\alpha$ for the set
II
}\label{alpha-tc}
\end{figure}

\begin{figure}[ht]
\epsfxsize=7cm
\begin{minipage}{0.4\textwidth}
\includegraphics[width=7cm,height=7.5cm,angle=-90]{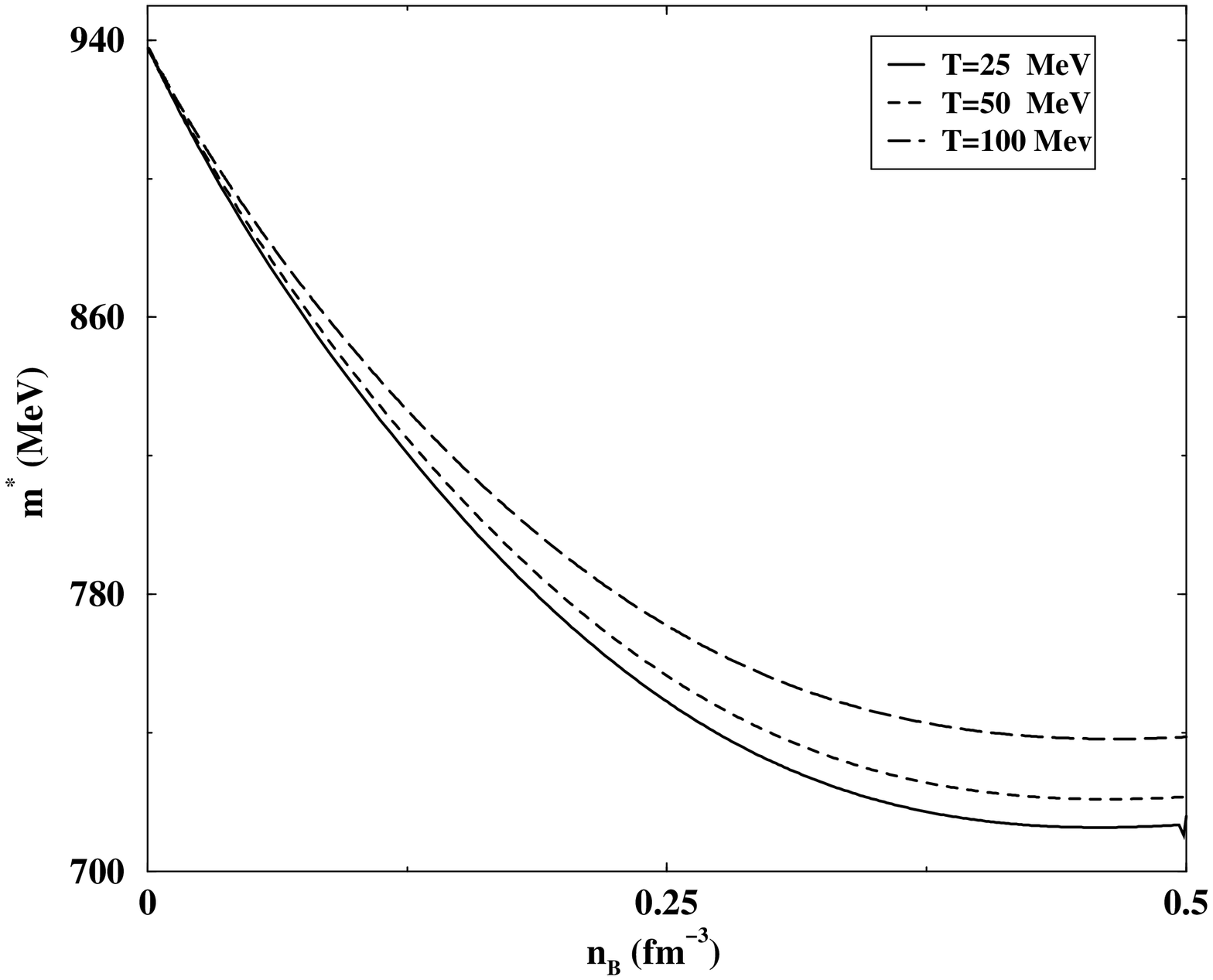}
\end{minipage} \
\begin{minipage}{0.4\textwidth}
\hspace*{1.5cm}
\includegraphics[width=7cm,height=7.5cm,angle=-90]{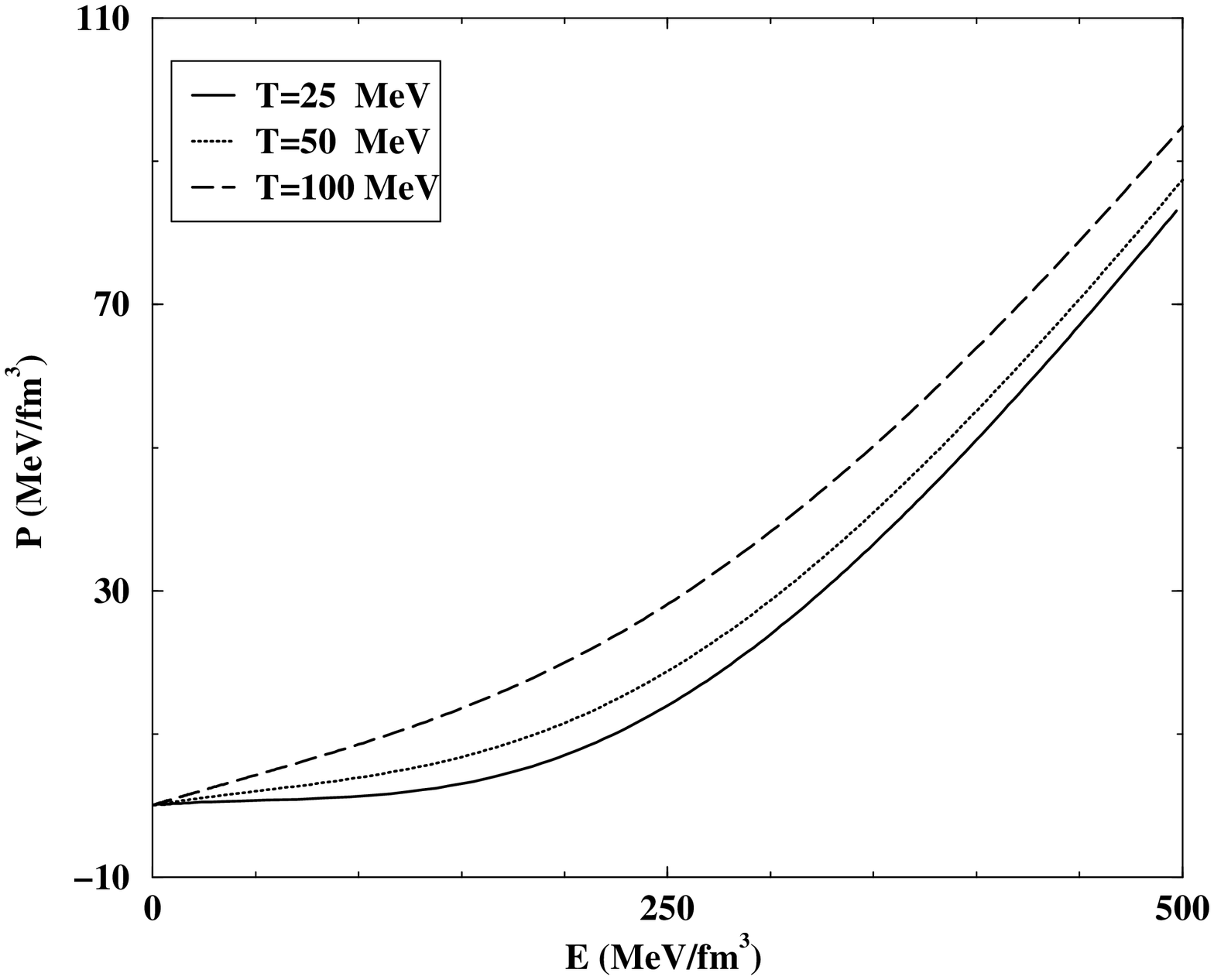}
\end{minipage}
\caption{The left panel shows effective mass versus baryon density at 
$T=$ 25, 50 and 100 MeV and the pressure as a function of energy is
displayed in the right panel for the same temperature.
}\label{mass-temp}
\end{figure}

\section{Summary and conclusions}

We presented a microscopic calculation of EOS in 
a relativistic framework based on the modified SU(2) chiral sigma
model with different parameter sets. 
In this model, we adopted an approach in which the mass of the isoscalar vector
field is generated dynamically. To ensure the empirical value of
incompressibility at saturation, we added higher-order terms of scalar meson field.
Thus the nucleon effective mass acquires a self-consistent density dependence 
on the scalar and vector meson fields.
Based on this model, we studied the effect of incompressibility ($K=210-380$ MeV) and 
effective mass($m^{\star}/m=0.8-9 $) on EOS near three
times the nuclear matter density. We compared our results with the
realistic EOS, prediction at low density region and also with the recently extracted
EOS from the heavy-ion collisions\cite{data02} at high density. The EOS 
obtained by NCS models overall agreed well with the experimental data and realistic
potential DBHF model. Among these NCS models, we found that the sets I and II are 
in agreement with the predicted EOS. In our discussions, we considered set II EOS for
the analysis of warm asymmetric nuclear system. The reason is
that it is compatible with recent heavy-ion collisions data\cite{sahu02}.
The change of effective mass is discussed as a function of the baryon density and temperature. We
found that it decreased with density around two times nuclear matter density and 
the variation was slow thereafter.

The liquid gas phase transition is studied within NCS model(set I, II and III). We 
found that the critical temperatures are 14.2, 16.8 and 20.4 MeV. The corresponding
$P_c$ and $n_c$ are 0.14, 0.22 and 0.36 MeV fm$^{-3}$ and 0.035, 0.044 and 
0.051 fm$^{-3}$, respectively. These are in the range of recent experimental
observation, $13.1\pm0.6$\cite{li94}. Precisely, set I is close to this value.
The binding energy per particle is also discussed with various temperature up to
25 MeV. We observed that the system becomes less bound with increasing temperature.

The EOS is also shown with variation of asymmetric parameter, $\alpha$ for a fixed temperature,
with and without $\rho-$meson contribution. The critical point decreased due to 
increase of $\alpha$. With inclusion of $\rho-$meson 
along with $\alpha$, it is reduced further.
This decrease in critical temperature in presence of $\rho-$meson is because of the 
strong repulsive force. This model worked well at low density region such as liquid
gas phase transition. The success of this model could be revisited for the study of
finite nuclei and at extreme densities and temperature. Works\cite{work03} are in 
progress to verify the validity of this model.

\section*{Acknowledgment}
 
We are thankful to P. Arumugam for useful discussion. One of us (TKJ) would like to
thank Institute of Physics, Bhubaneswar for the facilities provided to carry out
this work.

\end{document}